\documentclass[runningheads]{llncs}
\usepackage{graphicx}
\usepackage{times}
\usepackage{epsfig}
\usepackage{graphicx}
\usepackage{amsmath}
\usepackage{amssymb}
\usepackage{amsmath,bm}
\usepackage{mathrsfs}
\usepackage{booktabs}
\usepackage{float}
\usepackage{color}
\usepackage{hyperref}
\usepackage{url}

\begin{document}
\title{An Elastic Interaction-Based Loss Function for Medical Image Segmentation}
%
%
\author{Yuan Lan \and
Yang Xiang \and
Luchan Zhang}

\authorrunning{Y. Lan et al.}
%
\institute{Department of Mathematics,The Hong Kong University of Science and Technology,\\ Hong Kong SAR, China\\
\email{ylanaa@connect.ust.hk, \{maxiang,malczhang\}@ust.hk}\\}
\maketitle              
\begin{abstract}
Deep learning techniques have shown their success in medical image segmentation since they are easy to manipulate and robust to various types of datasets. The commonly used loss functions in the deep segmentation task are pixel-wise loss functions. This results in a bottleneck for these models to achieve high precision for complicated structures in biomedical images. For example, the predicted small blood vessels in retinal images are often disconnected or even missed under the supervision of the pixel-wise losses. This paper addresses this problem by introducing a long-range elastic interaction-based training strategy. In this strategy, convolutional neural network (CNN) learns the target region under the guidance of the elastic interaction energy between the boundary of the predicted region and that of the actual object. Under the supervision of the proposed loss, the boundary of the predicted region is attracted strongly by the object boundary and tends to stay connected. Experimental results show that our method is able to achieve  considerable improvements compared to commonly used pixel-wise loss functions (cross entropy and dice loss) and other recent loss functions on three retinal vessel segmentation datasets, DRIVE, STARE and CHASEDB1. The implementation is available at  \url{https://github.com/charrywhite/elastic\_interaction\_based\_loss}

\keywords{Medical Image Segmentation  \and Deep Learning\and Elastic Interaction}
\end{abstract}
\section{Introduction}
The medical image segmentation task is to automatically extract the region of interest, such as organs and lesions in medical images. Segmentation plays a critical role in medical image analysis since the information of the segmented region (such as length, width, angles) can be used for further diagnosis and treatment of related diseases. How to locate the region of interest in an exact manner is a major challenge.

In recent years, convolutional neural networks (CNNs) have shown ground-breaking effects to medical image segmentation task and different approaches have been proposed~\cite{medical1,medical2,medical3,medical4}. The widely adopted choice of the loss function in CNN methods is pixel-wise loss functions such as cross entropy. Pixel-wise loss functions classify each pixel of an image individually by comparing the corresponding ground truth and learning each pixel in an image equally. As a result, this type of loss functions are not necessarily robust in detecting the details and small structures which are critical in medical image analysis. For instance, in the retinal vessel segmentation task, the long, thin and weak vessels can be disconnected easily or even missed under the supervision of a pixel-wise loss function.

Before the flourishing of CNNs, variational image segmentation approaches \cite{variation2,variation,actour7}  based on intensity or textural information were  widely used  in dealing with the segmentation task.
 Among them, the active contour methods define an evolving curve under the force derived by some energy functional, and this evolving curve moves to minimize this energy. Different energy functionals have been proposed in the active contour methods, such as the balloon force model~\cite{actour2}, region-based methods~\cite{actour1,actour4}, gradient vector flow method (GVF)~\cite{actour3} and elastic interaction model~\cite{actour5,actour6}. Compared to the pixel-wise loss functions in CNN models, the curve evolution based on energy functionals in the active contour methods considers the topological or physical properties of the target objects rather than learning each pixel individually. However, these methods can only process one image at a time and cannot handle massive images in a collective way as in CNN methods.
 
  Recently, methods have been proposed \cite{loss1,loss2,loss3} to train a CNN by minimizing curve-evolving energy functionals. In \cite{loss1} and \cite{loss2},  the image segmentation task was represented as a minimization problem of the Mumford-Shah functional. While in \cite{loss3}, an objective minimization function based on boundary integral was defined. Despite these efforts, developing robust medical image segmentation methods, especially for long, thin structures in medical images, is still a challenging task.

In this paper, we propose an elastic interaction-based loss function  in a deep learning framework specifically for the long, thin structures that are commonly seen in medical images.
Under the supervision of this loss function, the predicted region will be attracted strongly by the object boundary through a global elastic interaction. Moreover, the self-interaction of the boundary of the predicted region will tend to smooth its own shape and will keep the predicted boundary connected. Due to these properties, this new loss function shows great advantages in tackling the
segmentation of thin, long structures, such as vessels, in medical images.
Also, one of the difficulties for most of the energy functional-based loss functions during training is the unstable performance in early steps, because the random initial guess in CNN may cause instabilities. Due to the long range nature, the minimization of the elastic interaction energy functional is not sensitive to the initialization. As a result, the proposed new CNN method with this elastic interaction-based loss function has demonstrated stable performance in the early stage of the training, as shown by the results of experiments in Section \ref{sec4}.

The proposed elastic interaction-based loss function can be implemented efficiently using the Fast Fourier Transform (FFT), making it easy to incorporate in CNN. We examine our new loss function on three retinal image datasets, DRIVE, STARE and CHASEDB1, and the results indicate that our method is able to achieve a considerable improvement compared to commonly used pixel-wise loss functions (cross entropy and dice loss) and other recent loss functions.


\section{Methodology} \label{sec3}
\subsection{Review of the framework of Deep Neural Network} 
In this section, we will briefly review the framework of deep neural network (DNN) under supervised learning. In supervised learning, the objective function (i.e. the loss function ) that we want to minimize is a function between the output of DNN and ground truth. The output of DNN is generated by forward propagation of the input, i.e. the input of DNN is processed by each hidden layer and activation layer in this network in the forward direction. Then the parameters within each hidden layer are updated by gradient descent in back propagation. The entire process of DNN is as follows,
\begin{equation}
\left\{
             \begin{array}{lr} 
             \min\limits_\theta \,\,\, \mathcal{L}\left(\theta\right) = \sum\limits_i f\left(u_N^{\left(i\right)}, u_{gt}^{\left(i\right)}\right) &  \\
            \text{Forward inference:} \, \, u_N^{\left(i,t\right)}\left(\theta\right) = \bold{W}_N^{\left(t\right)}\bold{\sigma}\circ\left(\bold{W}_{N-1}^{\left(t\right)}...\bold{\sigma}\circ\left(\bold{W}^{\left(t\right)}_{2}\bold{\sigma}\circ\left(\bold{W}_{1}^{\left(t\right)}u_0^{\left(i\right)}\right)\right)\right)\\
            \text{Back propagation:} \bold{W}_k^{t+1} = \bold{W}_k^{t} - \lambda\cdot \frac{\partial\mathcal{L}}{\partial u_N}\cdot\frac{\partial u_N}{\partial \bold{W_k}}
             \end{array}
\right.
\end{equation}
where $\mathcal{L}$ is the loss function to be minimized. $u_0,u_N,u_{gt}$ are input and output of network and ground truth, respectively, $i$ denotes the number of data and $t$ is the number of iterations,  $\theta = \left\{\bold{W_1},\bold{W_1},\dots,\bold{W_N}\right\}$ are the parameters to be trained, $\sigma$ is nonlinear scalar function, $\circ$ means element-wise operation, and $\lambda$ is step size in gradient descent.
In this work, we will proposed a new objective function for this DNN process.

\subsection{Elastic Interaction-Based Loss Function}

Our elastic interaction-based loss function is inspired by the elastic interaction between dislocations, which are line defects in crystals~\cite{dislocation1}. These dislocations are connected lines and the elastic interaction between them is long-range.
Inspired by this interaction energy, in deep learning segmentation task, we consider the boundary of the ground truth and that of the prediction as two curves with their own energies, during training process, the evolution of the boundary of predicted region  is under the supervision of their elastic interaction energy.

 Consider a parameterized curve $\gamma(s)$ in the $xy$ plane, which represents the boundary of a region. The curve has an orientation that could be either clockwise or counterclockwise.  The elastic energy of the curve  $\gamma$ is
 \begin{equation}\label{eqn:energy}
 	E = \frac{1}{8\pi}\int_\gamma\int_{\gamma'}\frac{d\bm{l}\cdot d\bm{l'}}{r},
 \end{equation}
 where $\gamma'$ is the curve $\gamma$ with another parameter $s'$, vector $d\bm{l}=\pmb\tau dl$ with $\pmb \tau$ being the unit tangent vector and
   $dl$ being the line element of the curve $\gamma$, and same for $d\bm{l}'$ of $\gamma'$,
   $\bm{r} =(x-x',y-y')$  is the vector between the points $(x,y)$ on $\gamma$ and $(x',y')$ on $\gamma'$, and the distance between these two points is
$r=\sqrt{(x-x')^2+(y-y')^2}$. This is the elastic energy stored in three dimensional space of a dislocation line $\gamma$ in the $xy$ plane. (Note that this  is a simplified version of the dislocation interaction energy~\cite{dislocation1}. Prefactor of the energy has also been omitted in this formulation.)

The notation $\gamma$ in Eq.~\eqref{eqn:energy} can also be understood as a collection of curves.
 Especially, for a system of two curves $\gamma_1$ and $\gamma_2$,  the total elastic energy is
 \begin{equation}\label{eqn:energy2}
 \begin{aligned}
 	E &= \frac{1}{8\pi}\int_{\gamma_1\cup\gamma_2}\int_{\gamma_1'\cup\gamma_2'}\frac{d\bm{l}\cdot d\bm{l'}}{r} \\
 	&=  \frac{1}{8\pi}\int_{\gamma_1}\int_{\gamma_1'}\frac{d\bm{l_1}\cdot d\bm{l_1'}}{r}+\frac{1}{8\pi}\int_{\gamma_2}\int_{\gamma_2'}\frac{d\bm{l_2}\cdot d\bm{l_2'}}{r}+\frac{1}{4\pi}\int_{\gamma_1}\int_{\gamma_2}\frac{d\bm{l_1}\cdot d\bm{l_2}}{r}.
 	\end{aligned}
 \end{equation}
In this expression, the first two terms are the self-energies of the two curves, respectively, and the last term is the interaction energy between them. Note that this total energy of two curves $\gamma_1$ and $\gamma_2$ depend on their orientations.

For a system of two coincident curves with opposite orientations,  i.e., $\gamma_2=-\gamma_1$, the total energy in Eq.~\eqref{eqn:energy2} vanishes. This means a perfect segmentation result if curve $\gamma_1$ is the boundary of the object (the ground truth) and $\gamma_2$ is the evolving boundary of prediction~\cite{actour5}.


Using an artificial time $t$, we describe the evolution of a curve $\gamma$ by minimizing  the total energy in the steepest descent direction, which is $\gamma_t=\bm{f}=-\frac{\delta E}{\delta \gamma}$, where $\gamma_t$ is the velocity of the curve $\gamma$ and $\bm{f}$ is the force on it. (This is essentially the dynamics of dislocations with mobility $1$ in its unit.) The obtained velocity of the curve $\gamma$ is
 \begin{equation}\label{eqn:gamma_t}
 	\gamma_t(x,y) = \bm{f}=\left(  -\frac{1}{4\pi}\int_\gamma\frac{\bm{r}\times d\bm{l}}{r^3} \right)\times \tau
= \left(  \frac{1}{4\pi}\int_\gamma\frac{\bm{r}\cdot \bm{n}_\gamma}{r^3} dl\right)\bm{n},
 \end{equation}
 where
$\bm{r} = \left(x-x\left(s\right),y-y\left(s\right)\right)$ is the vector between the point $\left(x,y\right)$ and a point $\left(x\left(s\right),y\left(s\right)\right)$ on $\gamma$, and $r = \sqrt{\left(x-x\left(s\right)\right)^2+\left(y-y\left(s\right)\right)^2}$ is the distance between the two points.
$\pmb \tau$ and $\bm{n}$ are the unit tangent vector and unit normal vector of $\gamma$ at the point $(x,y)$, respectively, and $\pmb \tau_\gamma$ and $\bm{n}_\gamma$ are the unit tangent vector and unit normal vector of $\gamma$ at the point $(x(s),y(s))$. In the derivation of this equation, we have used  $d\bm{l}=\pmb \tau_\gamma dl$, $\hat{\bm{z}}\times \pmb \tau=\bm{n}$, $\hat{\bm{z}}\times \pmb \tau_\gamma=\bm{n}_\gamma$, where $\hat{\bm{z}}$ is the unit vector in the $+z$ direction. Note that in Eq.~\eqref{eqn:gamma_t}, $\gamma$ could be a collection of curves.

Therefore, in the image space, derived from \eqref{eqn:energy} by using
$\nabla G_t =-\delta (\gamma_1)\bm{n}$ for the ground truth $G_t$ whose boundary is $\gamma_1$, and $\nabla H(\phi)=-\delta (\gamma_2)\bm{n}$ for the moving curve, where $\delta (\gamma_1)$ and $\delta (\gamma_2)$ are regularized Dirac delta functions of curves $\gamma_1$ and $\gamma_2$, the proposed loss function in CNN is,
\begin{small}
\begin{equation}\label{eqn:lossfunc}
\boldmath
L_{en} = \dfrac{1}{8\pi}\int_{\Omega}\!dxdy\int_{\Omega}\dfrac{\nabla \left(G_t+\alpha H\left(\phi\right)\right)\left(x,y\right)\cdot\nabla \left(G_t+\alpha H\left(\phi\right)\right)\left(x',y'\right) }{r}dx'dy',
\unboldmath
\end{equation}
\end{small}
where  $\phi$ is the level set representation of the moving curve $\gamma_2$ \cite{osher}, and $\alpha$ is a hyperparameter. Here $H(\cdot)$ is a smoothing Heaviside function which controls the width of the contour by regularization parameter $\beta$,
\begin{equation}
    H\left(\phi\right)=
\left\{
    \begin{array}{lcr}
    0 \qquad & {\rm if} \, \phi \leq -\beta   \\
    \frac{1}{2}\left(\sin\left(\frac{\pi\phi}{2\beta}\right)+1\right) \qquad &{\rm if} \, -\beta < \phi < \beta  \\
    1 \qquad & {\rm if} \, \phi \geq \beta.
    \end{array}
\right.
\end{equation}
The parameter $\alpha$ and the Heaviside function $H\left(\cdot\right)$ control the stregth of the elastic energy generated by the moving curve $\gamma_2$ compared with that generated by the object boundary $\gamma_1$.
In CNN context, we applied HardTanh activation function whose range is $[0,1]$ instead of using the above $H(\cdot)$ for convenience. The level set representation $\phi$ can be computed by $Prob-0.5$, where $Prob$ is the softmax output of target class.

Similarly, we describe the velocity of curve $\gamma_2$ (the boundary of predicted region) derived from ~\eqref{eqn:gamma_t} as,
\begin{equation} \label{eqn:backprop}
\begin{aligned}
    \boldmath
    v = -\frac{\partial L_{en}}{\partial \phi} & = -\frac{1}{4\pi}\int_{R^2}\frac{\boldsymbol{r}\cdot\nabla \left(G_t+\alpha H\left(\phi\right)\right)\left(x',y'\right)}{r^3}dx'dy'
    \unboldmath
\end{aligned}
\end{equation}
The velocity of curve $\gamma_2$ is the gradient of loss $L_{en}$ with respect to the level representation of predicted curve $\phi$, and it will be used in back propagation for training.

\subsection{Efficient Computation for Loss function and the Backward Gradient} \label{sec_time}
In order to compute the loss function ~\eqref{eqn:lossfunc} and gradient ~\eqref{eqn:backprop} efficiently in CNN, we reformulate them by Fast Fourier Transform which reduces the computational complexity from $O\left(N^2\right)$ to $O\left(N\log N \right)$.

Assume the Fourier transform of $\nabla \left(G_t+\alpha H\left(\phi\right)\right)$ in equation \eqref{eqn:lossfunc} is $d_{mn}$, where $m,n$ are the frequencies in Fourier space. It can be calculated that out the fourier transform of $\frac{1}{R} = \frac{1}{\sqrt{x^2+y^2}}$ is
$\widehat{\frac{1}{R}} = \frac{1}{2\pi\sqrt{m^2+n^2}}$.
Therefore, by Parseval's identity,   the loss function $L_{en}$ can be expressed in the Fourier space as
\begin{equation} \label{floss}
    \boldmath
    L_{en} = \sum\limits_{m,n} \sqrt{m^2+n^2}\cdot {|d_{mn}|}^2.
    \unboldmath
\end{equation}

From Eq.~\eqref{eqn:backprop},  the Fourier transform of the gradient of $L_{en}$ with respect to the output $\phi$
is
\begin{equation}
    \boldmath
    \widehat{\frac{\partial L_{en}}{\partial \phi}} = \frac{\sqrt{m^2+n^2}}{2} d_{mn}.
    \unboldmath
\end{equation}
According to this equation, we obtain the gradient for back propagation by inverse Fourier transform
\begin{equation}
    \boldmath
    \frac{\partial L_{en}}{\partial \phi} = \mathcal F^{-1}\left( \frac{\sqrt{m^2+n^2}}{2} d_{mn}\right)
    \unboldmath
\end{equation}
\subsection{Discussion on connectivity and fast convergence due to the strong
 long-range attractive interaction}

\begin{figure}[htbp]
    \centering
    \includegraphics[scale = 0.3]{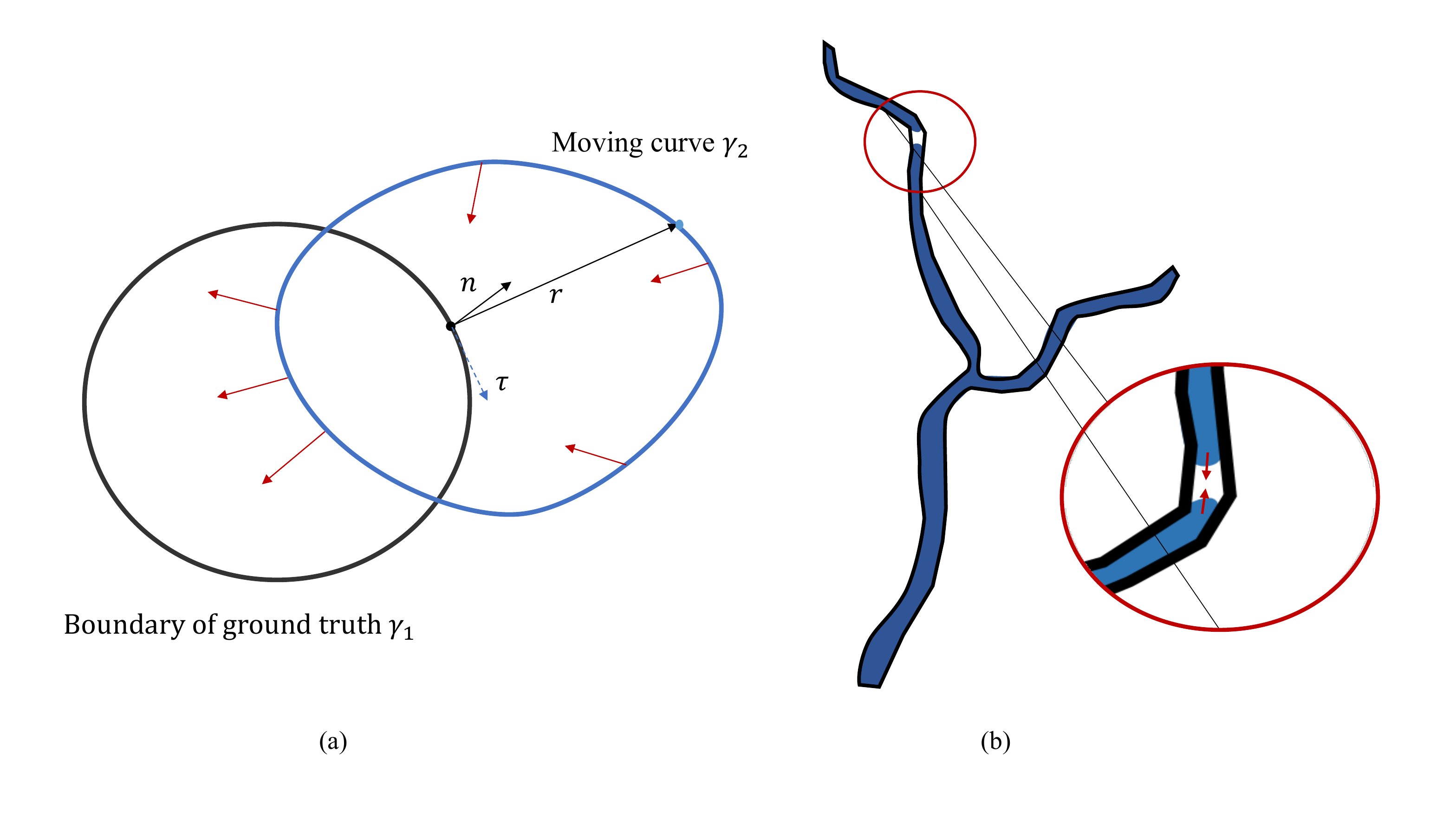}
    \caption{(a) Elastic interaction between the moving curve (the predicted boundary) $\gamma_2$ and the ground truth (object boundary) $\gamma_1$. Arrows on the moving curve $\gamma_2$ show schematically the directions of the interaction force acting on the moving curve $\gamma_2$.
    (b) Schematic illustration of recombination of a disconnected moving curve under elastic interaction in vessel segmentation. The black contour represents the true vessel, and the blue region represents the prediction. The boundaries of the disconnected parts of the prediction will be attracted to each other and recombine under the elastic interaction. Red arrows show the directions of the interaction force.}
    \label{fig:elastic}
\end{figure}

The elastic interaction  between two curves
 $\gamma_1$ and $\gamma_2$ is strongly attractive when the two curves have opposite orientations, as can be seen from the interaction energy given by the last term in Eq.~\eqref{eqn:energy2}. When the moving curve $\gamma_2$ (the predicted boundary) is set to have an opposite orientation with that of the object boundary $\gamma_1$ (ground truth), the moving curve will be attracted to the object boundary to minimize the total energy in Eq.~\eqref{eqn:energy2}, see Fig.~\ref{fig:elastic}(a); and when the two curves coincide, the energy minimum state is reached and the object is identified perfectly by the moving curve.

  This interaction between two curves is long-range because the energy density is inversely proportional to the distance between two respective points on them, which decays very slowly as the distance approaches to infinity, see the last term in Eq.~\eqref{eqn:energy2}.

  It can be shown that  near a curve $\gamma_1$, the interaction force experienced by a point on another curve $\gamma_2$ given in Eq.~\eqref{eqn:gamma_t} has the asymptotic property \cite{dislocation1,actour5}
  \begin{equation}
  \bm{f}\sim \frac{1}{r},
  \end{equation}
   where $r$ is the distance from the point on curve $\gamma_2$ to the curve $\gamma_1$. Therefore, the elastic interaction provides a strong attractive force to recombine a disconnected moving curve. See a schematic illustration in vessel segmentation in Fig.~\ref{fig:elastic}(b). (Here $\gamma_1$ and $\gamma_2$ are the two branches of the moving curve.)
   
    Moreover, the force generated by the moving curve on itself is proportional to $ \kappa\log\frac{1}{\beta}$, where $\kappa$ is the curvature of the curve, and
    $\beta$ is the half width of  the smooth Heaviside function $H(\cdot)$ \cite{dislocation1,actour5}. This self-force has the effect to smooth the moving curve.


\section{Experiment} \label{sec4}
%
\subsection{Dataset}
\noindent\textbf{DRIVE:} The Digital Retinal Images for Vessel Extraction (DRIVE) dataset \cite{drive} is provided by a diabetic retinopathy screening program in the Netherlands. This dataset includes 40 images ($3\times565\times584$ pixels) which are divided into a training set and test set, both containing 20 images and the corresponding ground truth. In our experiment, we split 5 images from training set as validation set. We train the DNN for vessel segmentation on 15 training data and evaluate our algorithm with 20 test images.

 \noindent\textbf{STARE:} STARE dataset ~\cite{stare1,stare2} consists of 20 color retinal images with size $700\times 605$ pixels. 15 images were used for training  and the remaining were used for validation. The hand labeled ground truth are provided for all images in this dataset.

 \noindent\textbf{CHASEDB1:} This dataset \cite{chasedb1} includes 30 color images with size $999\times 960$ pixels. We split 5 images for validation, 25 images for training. The ground truth is provided for each image. 

\subsection{Implementation Details} \label{sec_implement}

\noindent\textbf{Data preprocessing:} Since the retinal images are low contrast usually, we selected the green channels of images for training because the green channels have higher contrast than the red and blue channels. Then all values in images are normalized into $[0,1]$.

\noindent\textbf{Architecture and training:}
In our experiment, we employed U-Net \cite{medical3} as the end-to-end convolutional neural network architecture on our experiment. Based on this architecture, we trained our model 50 epochs with a batch size equal to 6 on training dataset. Adam optimizer was used to optimize this model with learning rate of 0.001. For our elastic interaction loss function in Eq.~\eqref{eqn:lossfunc}, we set hyperparameter $\alpha = 0.35$, and $\beta = 0.25$ in the Hardtanh (smoothing Heaviside) function.
We implemented our model based on Pytorch \cite{pytorch} and ran all the experiments by the machine equipped with an TESLA P100 with 16GBs of memory.

\noindent\textbf{Evaluation metrics:} To evaluate the performance of segmentation, sensitivity, specificity, f1 score and AUC (the area under the receiver operating characteristics curve) were calculated.

\subsection{Results}
\begin{table}
\centering
\caption{Results of the proposed elastic loss function (EL) and other loss functions on DRIVE, STARE and CHASEB1. Sen, Spec, F1 score, AUC in the table are Sensitivity, Specificity, F1 score and the area under the receiver operating characteristics curve, respectively. } \label{form:testresult}
\setlength{\tabcolsep}{4.5mm}{
\begin{tabular}{lccccr}
\hline
Dataset&Method& Sen & Spec&F1 score&AUC \\
\hline
&CE& 0.9667& 0.7721& 0.8025&0.8694\\
&DICE& 0.9615& 0.7287& 0.7658& 0.8451\\
DRIVE&AC~\cite{loss1}& 0.9647& 0.7682& 0.7921& 	0.8664\\
&SL~\cite{loss3}& 0.9660& 0.7940& 0.8043& 0.8800\\
&\textbf{Proposed EL}& \textbf{0.9664}& \textbf{0.8067}&\textbf{0.8093}& \textbf{0.8866}\\
\specialrule{0.05em}{3pt}{3pt}
&CE& 0.9564& 0.7296&0.7214&0.8430\\
&DICE& 0.9403&0.6588& 0.6212& 0.7996\\
STARE&AC~\cite{loss1}& 0.9419& 0.6461& 0.6144& 0.7940\\
&SL~\cite{loss3}& 0.9540& 0.7434&0.7146& 0.8487\\
&\textbf{Proposed EL}& \textbf{0.9576}& \textbf{0.7449}& \textbf{0.7304}& \textbf{0.8513}\\
\specialrule{0.05em}{3pt}{3pt}
&CE& 0.9526& 0.8408& 0.8245& 0.8967 \\
&DICE& 0.9276&  0.7243& 0.7186& 0.8259\\
CHASEDB1&AC~\cite{loss1}& 0.9506& 0.8215& 0.8145&  0.8861\\
&SL~\cite{loss3}& \textbf{0.9545} & 0.8207& \textbf{0.8258}& 0.8876\\
&\textbf{Proposed EL}& 0.9526&\textbf{0.8428}& 0.8248&  \textbf{0.8977}\\
\hline
\end{tabular}}
\label{tab:quanti}
\end{table}

\noindent\textbf{Quantitative evaluation:} 
We evaluated our model with the elastic interaction-based loss function (EL) on three datasets comparing to models with cross entropy loss function (CE), the dice coefficient loss (DICE), active contour loss (AC)~\cite{loss1} and surface loss~\cite{loss3}. The quantitative results of our proposed loss on three datasets are shown in Table~\ref{tab:quanti}. The experimental results show that our proposed loss function has better performance than other loss functions by achieving highest Sensitivity, Specificity, F1 scores and AUC on DRIVE and STARE, highest AUC and Specificity on CHASEDB1.

\begin{figure}[hbtp]
\begin{center}
  \includegraphics[scale = 0.45]{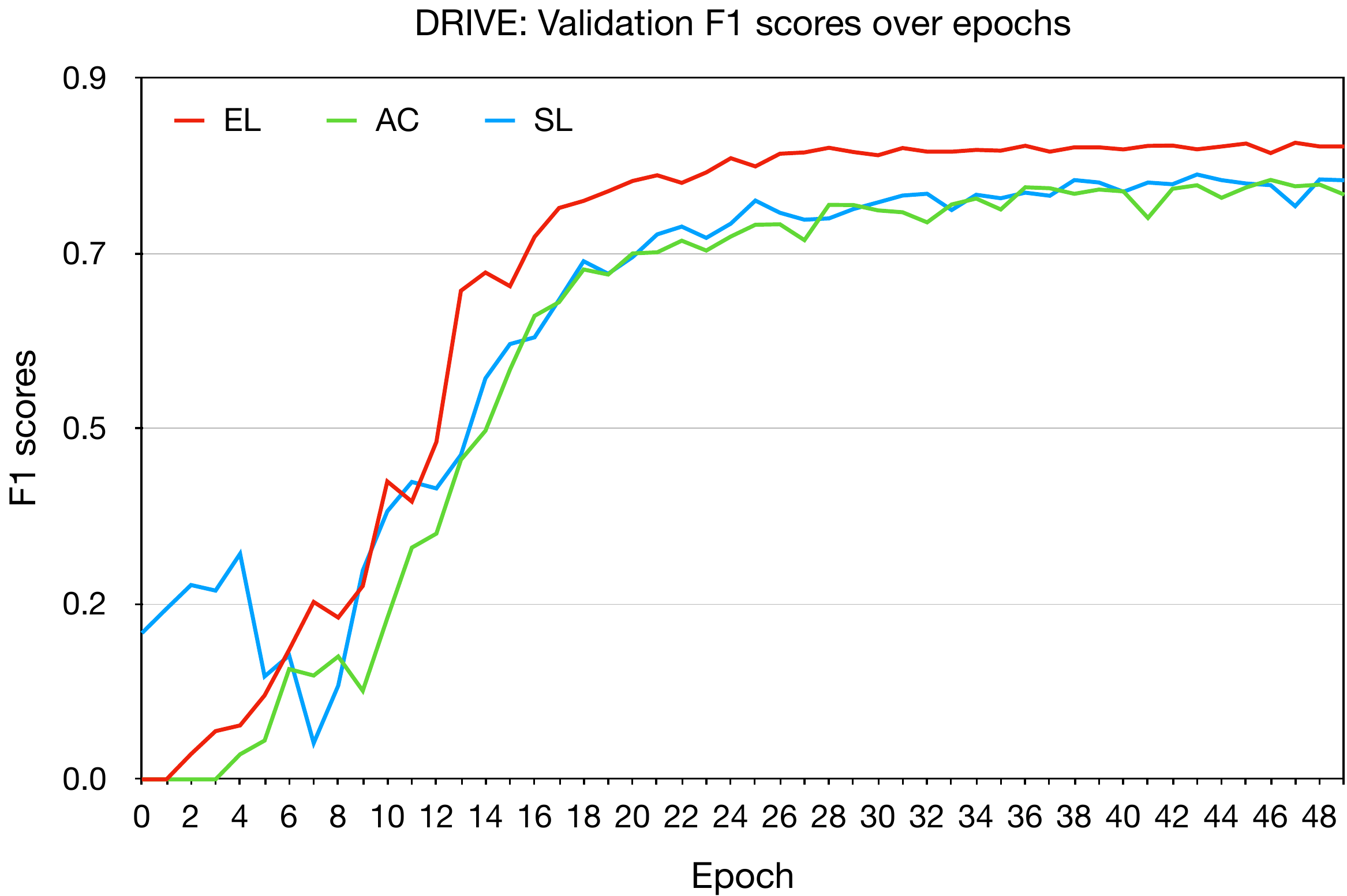}
  \caption{Validation F1 scores durning training steps on DRIVE dataset with three energy functional-based loss functions. EL: red curve, AC: green curve and SL: blue curve. }\label{fig:ac}
  \end{center}
\end{figure}

\begin{figure}[hbt] 
\begin{center}
  \includegraphics[scale = 0.45]{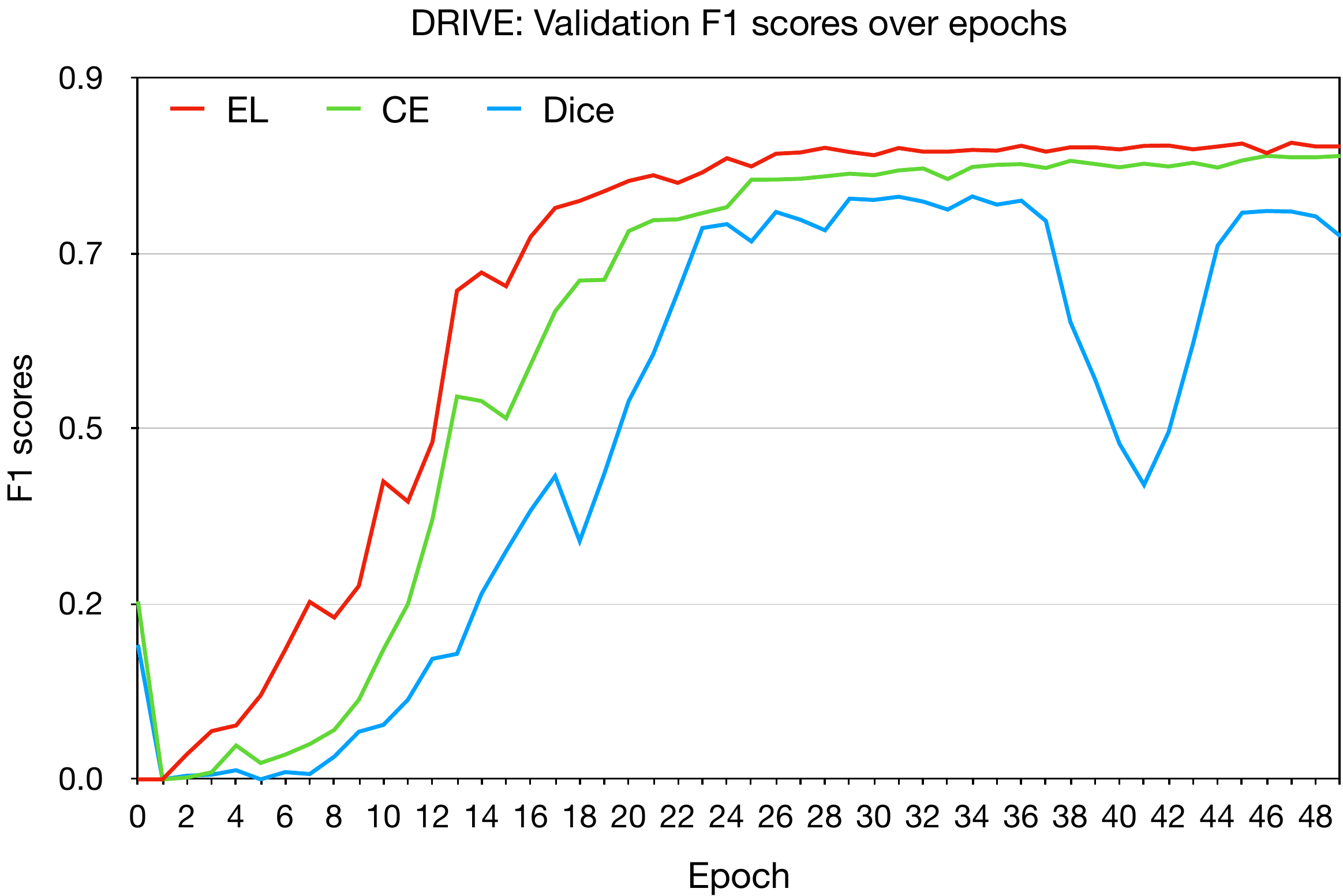}
  \caption{Validation F1 scores durning training steps on DRIVE dataset with EL and other two traditional loss functions CE and Dice. EL: red curve, CE: green curve and Dice: blue curve.} \label{fig:pixel}
  \end{center}
\end{figure}

We examine the stability and convergence of Deep neural network in training steps with our proposed elastic interaction-based loss (EL) compared with those of traditional loss functions (CE and Dice) and other energy functional-based loss functions (AC and SL). In Fig.\ref{fig:ac}, we can observe that the performance of network under the guidance of EL in early training steps is more stable than that of other two energy functional-based loss functions, and it also shows higher F1 scores compared with those of other two loss functions. Similarly, in Fig.\ref{fig:pixel}, compared to performance of traditional loss functions, EL shows more stable training performance, faster convergence and better result.

\begin{figure}[htbp]
    \centering
    \includegraphics[width=4.5in]{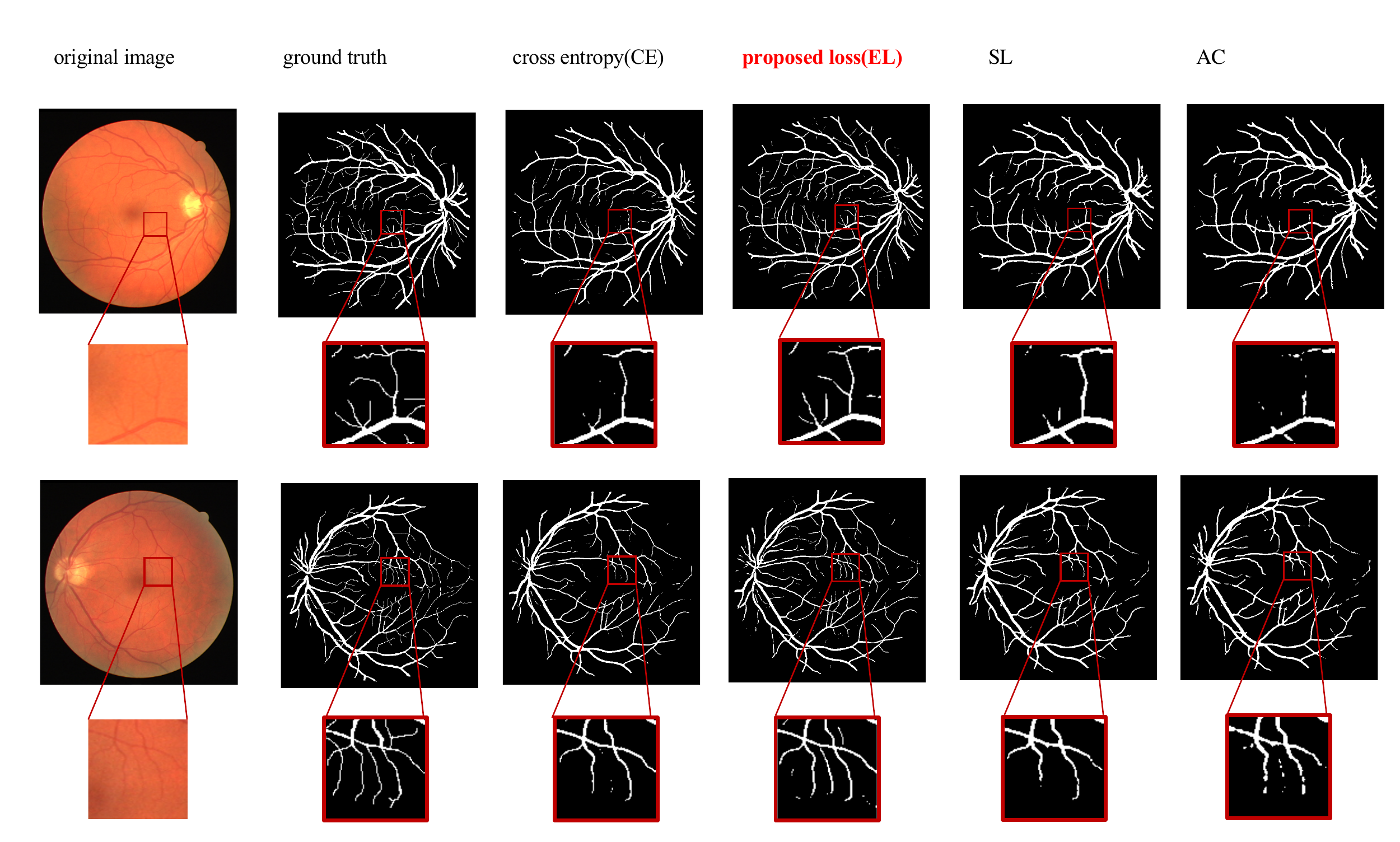}
    \caption{The example test results obtained by our proposed loss and other loss functions. From left to right, original images, ground truth, segmentation result by cross entropy (CE), our proposed loss (EL), surface loss (SL),active contour loss (AC).}
    \label{fig:drive}
\end{figure}

\noindent\textbf{Qualitative evaluation:}
Figure \ref{fig:drive} shows the qualitative results of different loss functions. From Fig.\ref{fig:drive}, we can see that the results under the proposed loss outperform  other loss functions because our method can segment the much more details of vessel among all of the mentioned loss functions. Specifically, in the first row of Fig.\ref{fig:drive}, the result under the supervision of cross entropy can not extract the thin and long vessel within red box and the prediction is disconnected, thus this segmentation lost a lot of details. While in the prediction from our proposed loss (EL), we can see that it segments the full vessels successfully and the predicted vessels are connected. Therefore, we conclude that the elastic interaction loss function has better performance than other three loss functions in extracting details of images (especially for thin and long structures). Same for the second row.

\noindent\textbf{Computational time:}
As we mentioned in Section \ref{sec_time}, computational complexity of our proposed elastic loss (EL) is 
$O\left(N\log N \right)$. In practice, running time of EL durning the training and test processes on our experimental device mentioned on Section \ref{sec_implement} is comparable with those of traditional loss functions. For DRIVE dataset, it took \textit{$297$s} on training and took \textit{$5.5$s} to segment one test image. CE and DICE took \textit{$331$s} and \textit{$300$s} on training, respectively and they use \textit{$5.5$s} on segmenting each test image.  While for STARE, the running time of EL (\textit{Training: $501$s, Test per image: $1.8$s}) is also very close to that of CE (\textit{Training: $539$s, Test per image: $1.6$s}) and DICE (\textit{Training: $511$s, Test per image: $1.6$s}). Those loss functions  showed similar behavior on CHASEDB1.

\section{Conclusions and Discussion}
In this paper, we propose a new elastic interaction-based loss function that can connect the whole segmented boundary strongly, thus it has great advantages in the segmentation of details of images in medical image datasets. Experimental results show that the proposed loss function indeed enhances the ability of CNN to segment medical images with complex structures and is able to achieve better performance comparing to the commonly used loss functions and other energy functional based loss functions.
 We would like to remark that the proposed loss function is developed to detect the details of long, thin structures such as vessels in medical images. When it applies to general images without these structures, the proposed method may not necessarily have obvious advantages.

\section*{Acknowledgement}
This work was supported by The Hong Kong University of Science and Technology IEG19SC04.
%
%
%
 \bibliographystyle{splncs04}
 \bibliography{export_2}

%
%
%
%
\end{document}